\documentclass[twocolumn,pra,showpacs,preprintnumbers,amsmath,amssymb]{revtex4}

\usepackage{graphicx}
\usepackage[usenames]{color}
\begin{document}

\title{Photon correlations in multi-mode waveguides}

\author{Eilon Poem}
\email{eilon.poem@weizmann.ac.il}
\affiliation{Department of Physics of Complex Systems, Weizmann Institute of Science, Rehovot 76100, Israel}
\author{Yaron Silberberg}
\affiliation{Department of Physics of Complex Systems, Weizmann Institute of Science, Rehovot 76100, Israel}

\date{\today}

\begin{abstract}
We consider the propagation of classical and non-classical light in multi-mode optical waveguides.
We focus on the evolution of the few-photon correlation functions, which,
much like the light-intensity distribution in such systems, evolve in a periodic manner, culminating in the `revival' of the initial correlation pattern at the end of each period.
It is found that when the input state possesses non trivial symmetries, the correlation revival period can be longer than that of the intensity, and thus the same intensity pattern can display different correlation patterns.
We experimentally demonstrate this effect for classical, pseudo-thermal light, and compare the results with the predictions for non-classical, quantum light.
\end{abstract}


\maketitle

Periodic recurrences, or `revivals', of an initial condition are a general phenomenon common to many physical systems~\cite{Robinett04}. It was experimentally observed for electrons in Rydberg atoms~\cite{Yeazell90}, atomic~\cite{Clauser94,Chapman95} and molecular~\cite{Brezger02} matter waves, Bose-Einstein condensates~\cite{Deng99}, and many more. In optics, this effect dates back to the observations of periodic imaging of diffraction gratings by Talbot in 1836~\cite{Talbot}, which were explained in terms of Fresnel diffraction by Rayleigh in 1881~\cite{Rayleigh}.
The common feature of all these systems is that their relevant eigen-modes (i.e. those initially populated) accumulate phase 
in rates that are all, either exactly or approximately, integer multiples of the same number. Thus, there exists a point 
at which the phases accumulated by the eigen-modes are all integer multiples of $2\pi$, and the initial wavefunction is reconstructed. This effect then repeats itself periodically. Furthermore, at rational fractions of the revival period, `fractional revivals' occur: the initial wavefunction reappears in multiple, shifted copies of itself.
In applied optics, these phenomena, collectively known as multi-mode interference (MMI), are utilized for multiple beam splitting and combining~\cite{Soldano95,Heaton99}. Recently~\cite{Peruzzo11}, such MMI-based multi-port integrated devices were shown to function as multiple beam splitters also at the single-photon level, displaying quantum interference effects.

The revival of the initial intensity pattern in optical systems through MMI is well described within wave optics, and do not require the quantization of the optical field.
Quantum-optical effects, such as Hanbury Brown and Twiss~\cite{HBT}, or Hong-Ou-Mandel interference~\cite{HOM87}, affecting the relative distribution of single photons, can be revealed only through measurement of photon-correlation functions.
Second-order correlation functions have been recently investigated for diffraction gratings~\cite{Lou09,Song10,Song11}, and it was shown that the recurrence period of the second order spatial correlation function is twice larger than that of the light intensity.
As MMI systems show periodical recurrences of the light intensity, much like diffraction gratings, and as they can be rather easily miniaturized and integrated in photonic circuits~\cite{Soldano95,Peruzzo11}, it is interesting to study the behavior of correlation functions in these systems. Two questions immediately arise: Do second-order correlations in MMI systems have a double period in respect to the total intensity, as in diffraction gratings? and how do higher-order correlations behave?

Here we theoretically and experimentally explore few-photon correlations in MMI systems. As a simple but rather flexible model experimental system we use a two-mirror planar multi-mode waveguide.

In this system, a general initial wavefunction recurs every $z_0=8D^2/\lambda$~\cite{Soldano95,Heaton99}, where $D$ is the distance between the mirrors and $\lambda$ is the wavelength of the incident light. As the intensity distribution and the correlation functions of all orders are functions of the wavefunction, they too recur at this distance. 
If the input state possesses only trivial symmetry, meaning that its wavefunction is an eigenstate only of the identity transformation, $z_0$ is the shortest recurrence period. Thus, for such a `completely asymmetric' initial wavefunction, the intensity and all correlations have the \emph{same} recurrence period, $z_0$.

\begin{figure}[tbh]
\centering
  \includegraphics[width=0.48\textwidth]{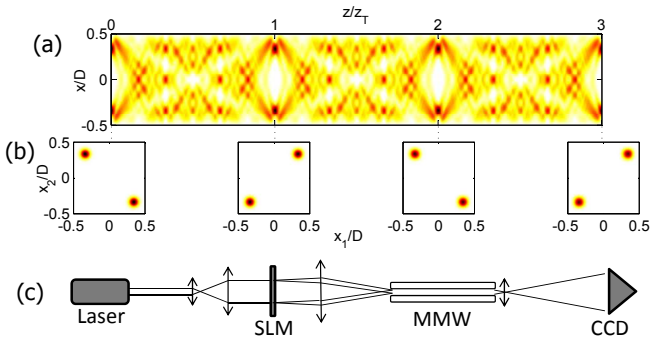}
  \caption{(a) Calculated intensity pattern for two beams propagating through a multimode waveguide of width $D$ along the z direction. Light (dark) color represents low (high) intensity. Periodic revivals of the initial intensity pattern are seen at integer multiples of the Talbot distance $z_T$. (b) Calculated correlation functions for two photons entering the waveguide one on each beam. Here the revival period is $2z_T$. (c) Schematics of the experimental system. SLM - spatial light modulator; MMW - multi-mode waveguide; CCD - charge coupled device.}
  \label{fig:1}
\end{figure}

This situation changes when the initial wavefunction has a non-trivial symmetry. 
Such a case is that of two uncorrelated, identical beams entering the waveguide at
locations symmetric about its center. This case applies to a quantum input state containing a single photon in each beam, as well as to two thermal input beams. The total intensity distribution along the waveguide for such initial states is composed of the \emph{incoherent} sum of the intensity distributions originating from each beam separately. Due to the initial symmetry, the two distributions are mirror images of one another. Thus, for all planes where these distributions are composed of at most two lobes, their sum always results in two identical lobes.
Most generally, such planes repeat every Talbot distance, \mbox{$z_T$=$z_0/4$}~\cite{Soldano95,Heaton99}, as can be seen in the numerical calculation presented in Fig.~\ref{fig:1}(a).

The only nontrivial transformation keeping an uncorrelated symmetric input state (either quantum or classical) unchanged is the symmetric exchange of photons between the two sides of the waveguide. Such an exchange occurs during the evolution through the waveguide every \mbox{$z_0/2$=$2z_T$}~\cite{Soldano95,Heaton99}. Thus, the second order correlation function, determined by the two-photon part of the state, recurs every $2z_T$. This period is twice larger than the recurrence period of the corresponding total intensity. Two different correlations patterns can thus exist for the same intensity distribution. For the separated quantum input state, as shown in the calculation presented in Fig.~\ref{fig:1}(b), the two photons can be either separated between the two beams, or bunched at the same beam.

This principle can be extended to any case of N uncorrelated identical beams entering the waveguide at the points $x_n=(2n-1-\mbox{N})D/2$N. As for the case of two uncorrelated symmetric beams discussed above, the only non-trivial symmetry of the N$^{\mbox{\small{th}}}$ order correlation function is in the symmetric exchange of photons between the two sides of the waveguide. The period of this correlation function is thus $2z_T$ for all N. However, the intensity distribution recurs every time a single input beam divides into up to N beams, a situation occurring every $z_T$/N~\cite{Heaton99}. These cases thus show a correlation recurrence period 2N times longer than that of the total intensity.

Experimentally, we demonstrate these effects up to the 3$^{\mbox{\small{rd}}}$ order correlation function for up to 3 beams of thermal light.
Our experimental setup is described in Fig.~\ref{fig:1}(c). Its main part is a waveguide made of two parallel planar metallic mirrors, placed on translation stages for adjusting the distance between them - the waveguide's width.
Two or three beams of pseudo-thermal light are focused on the input facet of the waveguide, and the two or three photon correlation function is measured at its output facet. This is done for different widths, $D$, controlling the recurrence distance, $z_0$, for the fixed waveguide length,~$L$.

The widths used in the experiment are between 57 and 74 $\mu$m. The wavelength of the incident light is 532 nm. The waveguide thus supports over one hundred modes. The incident light is focused on the entrance to the waveguide by a cylindrical lens. This is done in order to minimize the diffraction in the direction parallel to the mirrors. The focusing lens used was of NA$\approx$0.1, creating a spot about 5$\mu$m in width. Therefore, only the lowest $\sim$10\% of the waveguide modes are excited, and accurate MMI imaging is ensured~\cite{Soldano95}.
Due to the finite conduction of the mirrors, the waveguide is slightly birefringent. To avoid any complication that may arise, the incident light was polarized perpendicularly to the mirrors, and thus only the TM modes were excited.
In order to create multiple spots at the entrance to the waveguide, the incoming light was first passed through an amplitude and phase spatial light modulator (SLM) used as a variable transmission grating. A second cylindrical lens was placed after the waveguide in order to create a magnified image of its output facet on a CCD camera.

Thermal light was simulated by repeating the experiment for many different relative phases between the incident beams, controlled by the SLM. For each phase realization, the spatially resolved correlation function was produced from the recorded intensity, and all such single phase correlation functions were summed up to give the thermal light correlation function~\cite{Bromberg09}.
\begin{figure}[tbh]
\centering
  \includegraphics[width=0.48\textwidth]{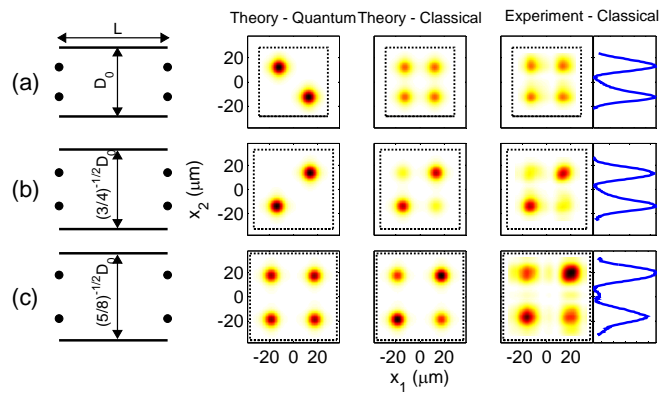}
  \caption{Calculated and measured spatial intensity correlation maps for two beams at symmetric input locations. The first panel on the left of each row shows a schematic illustration of the waveguide settings. Black circles denote the positions of the input (left) and output (right) beams. The second and third panels present calculated correlation maps for quantum and thermal input states, respectively. The dotted squares mark the locations of the mirrors. The fourth panel presents the  measured correlation map for pseudo-thermal input in a waveguide with $L$=4.85 cm, and the measured intensity distributions, for 3 waveguide settings. (a) Imaging,  $D$=$D_0$=$\sqrt{\lambda L/8}$=$57 \mu m$. (b) Equal two-way beam-splitting, $D$=$66 \mu m$. (c) Unequal two-way beam-splitting, $D$=$72 \mu m$. Here the effect is specific to input beams located at $\pm D/4$.}
  \label{fig:2}
\end{figure}

Fig.~\ref{fig:2} presents calculated and measured intensity correlation functions between two beams, introduced into the waveguide at symmetric locations, for three different waveguide settings, (a)-(c), schematically presented on the left panel in each row. The second panel in each row presents the correlation map of light at the output facet of the waveguide, calculated for a quantum input state in which there is exactly one photon in each beam: $\psi^q_{in}=|11\rangle$, where the left (right) number counts the photons on the upper (lower) beam.
The third (fourth) panel presents the calculated (measured) output correlation map for a classical, thermal input state, where all photon configurations have the same probability. The fourth panel presents also the measured total intensity distribution. The two-photon part of such the thermal state is given by the following density matrix, written in the basis $\{|20\rangle, |11\rangle, |02\rangle\}$:
\begin{equation}\label{eq:rho_in}
\rho^{th}_{in}=\frac{1}{4}\mbox{\small{$\left(
                                      \begin{array}{ccc}
                                        1 & 0 & 0 \\
                                        0 & 2 & 0 \\
                                        0 & 0 & 1 \\
                                      \end{array}
                                    \right)$}}.
\end{equation}
The factor of two for the $|11\rangle$ state is due to the indistinguishability of the photons, for which this state is reached from two configurations. Due to this factor and the lack of non-diagonal terms, thermal light shows correlations similar to those of the $|11\rangle$ quantum light, though always with a `background' coming from the $|20\rangle$ and $|02\rangle$ states~\cite{Bromberg09}.
%

In Fig~\ref{fig:2}(a) the width, $D$, of the waveguide is set such that its length, $L$, which is fixed at 4.85 cm, would be equal to the imaging length: $D$=$D_0$=$\sqrt{\lambda L/8}$. For $\lambda$=$532$~nm, $D_0$=$57 \mu$m. In this case, the field at the output facet is a complete reconstruction of the field at the input facet. This includes the reconstruction of all correlations. For a quantum two-photon input state, the photons at the output are also completely separated (`anti-bunched'), and the probability of finding two photons on one side (`bunching') stays zero, as seen in the calculated correlation map on the second panel, where the `diagonal' elements vanish. For two beams of classical, thermal light, the photons have the same probability to bunch or anti-bunch, as indeed seen in both the calculated and measured intensity correlation maps presented in the third and fourth panels, respectively.

In Fig.~\ref{fig:2}(b) the waveguide width was set to $D_0/\sqrt{3/4}$, meaning that \mbox{$L$=$3z_0/4$=$3z_T$}. At this distance, for input beams at symmetric locations, the waveguide acts as an equal two-way beam-splitter~\cite{Soldano95,Heaton99,Peruzzo11}. The quantum input state $|11\rangle$ would therefore be transformed into the path-entangled state,
\begin{equation}
\psi^q_{out}=\frac{1}{\sqrt{2}}(|20\rangle+|02\rangle).
\end{equation}
The probability for anti-bunching would now be zero. This is seen in the calculated correlation map shown on the second panel. Each component of the thermal state $\rho^{th}_{in}$ (Eq.~\ref{eq:rho_in}) will undergo the beam-splitter transformation by itself. The resulting density matrix is:
\begin{equation}
\rho^{th}_{out}=\frac{1}{8}\mbox{\small{$\left(
                                      \begin{array}{ccc}
                                        3 & 0 & 1 \\
                                        0 & 2 & 0 \\
                                        1 & 0 & 3 \\
                                      \end{array}
                                    \right)$}}.
\end{equation}
In this state it is three times more probable that the photons would bunch than that they would anti-bunch. This is readily seen also in the numerical calculation presented on the third panel, where this ratio is the ratio between the sum of all diagonal lobes and that of all non-diagonal lobes. The bunching-anti-bunching ratio extracted in the same way from the measured correlation map, presented in the fourth panel, is 2.5$\pm$0.1. The deviation from the theoretical value may be attributed to background coming from parasitic effects such as mode mixing induced by mirror imperfections.

Fig.~\ref{fig:2}(c) presents correlation functions for a width of $D$=$D_0/\sqrt{5/8}$ (\mbox{$L$=$5z_T/2$}). For input beams at $\pm D/4$ from the middle of the waveguide, this is the intermediate case between the cases presented in Fig.~\ref{fig:2}(a) and in Fig.~\ref{fig:2}(b). In this case, a single photon would split unequally between the two beams, with amplitudes proportional to $\cos\pi/8$ and $\sin\pi/8$~\cite{Heaton99}. 
The output state of two photons entering in the separated state is now an equal superposition of the separated state and the bunched, path-entangled state,
\begin{equation}
\psi^q_{out}=\frac{1}{\sqrt{2}}|11\rangle+\frac{1}{2}(|20\rangle+|02\rangle).
\end{equation}
Therefore, there is an equal probability for bunching and anti bunching. This is shown in the second panel of Fig.~\ref{fig:2}(c). The thermal state would be transformed by this unequal beam-splitter into the state:
\begin{equation}
\rho^{th}_{out}=\frac{1}{16}\mbox{\small{$\left(
                                      \begin{array}{ccc}
                                        5 & i\sqrt{2} & 1 \\
                                        -i\sqrt{2} & 6 & -i\sqrt{2} \\
                                        1 & i\sqrt{2} & 5 \\
                                      \end{array}
                                    \right)$}}.
\end{equation}
The probability ratio between bunching and anti-bunching is now 5/3. This is seen in both calculation and measurement, the measured ratio being 1.5$\pm$0.1.

It should be stressed that in all the cases presented in Fig.~\ref{fig:2}, the output intensity distributions are similar: they all include two identical lobes. The difference between the cases is revealed only in the correlation pattern, for a thermal as well as for a quantum input state. Note that for the special case of input beams at $\pm D/4$, the correlation recurrence period is \emph{four} times longer than that of the intensity.

We now turn to discuss experiments with three input beams. We consider the case of equally spaced, identical beams entering the waveguide at $x_1$=$-D/3$, $x_2$=0, and $x_3$=$D/3$. For this case, each one of the beams splits into three every $z_T/3$. Although each time this three-ways splitting is in different weights, the total intensity of all three beams together is the same for all these points. The different weights of the single beam splitting are manifested in the correlation functions, which, as discussed above, recur only every $2z_T$, a period \emph{six} times longer than that of the total intensity.

For example, while for a width $D$=$D_0/\sqrt{2/3}$ ($L$=$8z_T/3$), the waveguide functions as an equal three-way beam splitter, for a width $D_0/\sqrt{7/12}$ ($L$=$7z_T/3$) it functions as an un-equal three-way beam splitter~\cite{Heaton99}. While the total intensity distribution for these two cases is the same, featuring three equal lobes, the states, and hence the correlations, are different.
For a quantum state of the form $|111\rangle$ input into the equal three-way beam-splitter, the output wavefunction is,
\begin{equation}
\psi^q_{out}=\tfrac{\sqrt{2}}{3}e^{i\tfrac{2\pi}{3}}\left(|300\rangle-|030\rangle+|003\rangle\right)+\tfrac{1}{\sqrt{3}}|111\rangle.
\end{equation}
It carries mostly three photon bunching, along with some complete anti-bunching, but no two-photon bunching~\cite{Tichy10}.
In contrast, the unequal three-way beam-splitter produces the wavefunction,
\begin{eqnarray}
\nonumber \psi^q_{out}=\tfrac{1}{3\sqrt{2}}\left(|300\rangle+2|030\rangle+|003\rangle\right)\\
-\tfrac{1}{\sqrt{6}}\left(e^{i\tfrac{2\pi}{3}}|210\rangle+|201\rangle+|102\rangle+e^{i\tfrac{2\pi}{3}}|012\rangle\right),
\end{eqnarray}
which carries mostly two-photon bunching, some three photon bunching, and no complete anti-bunching. Since we would like to have a measure for three-photon bunching or anti-bunching, we calculate and measure the \emph{third} order intensity correlation function.
\begin{figure}[tbh]
\centering
  \includegraphics[width=0.48\textwidth]{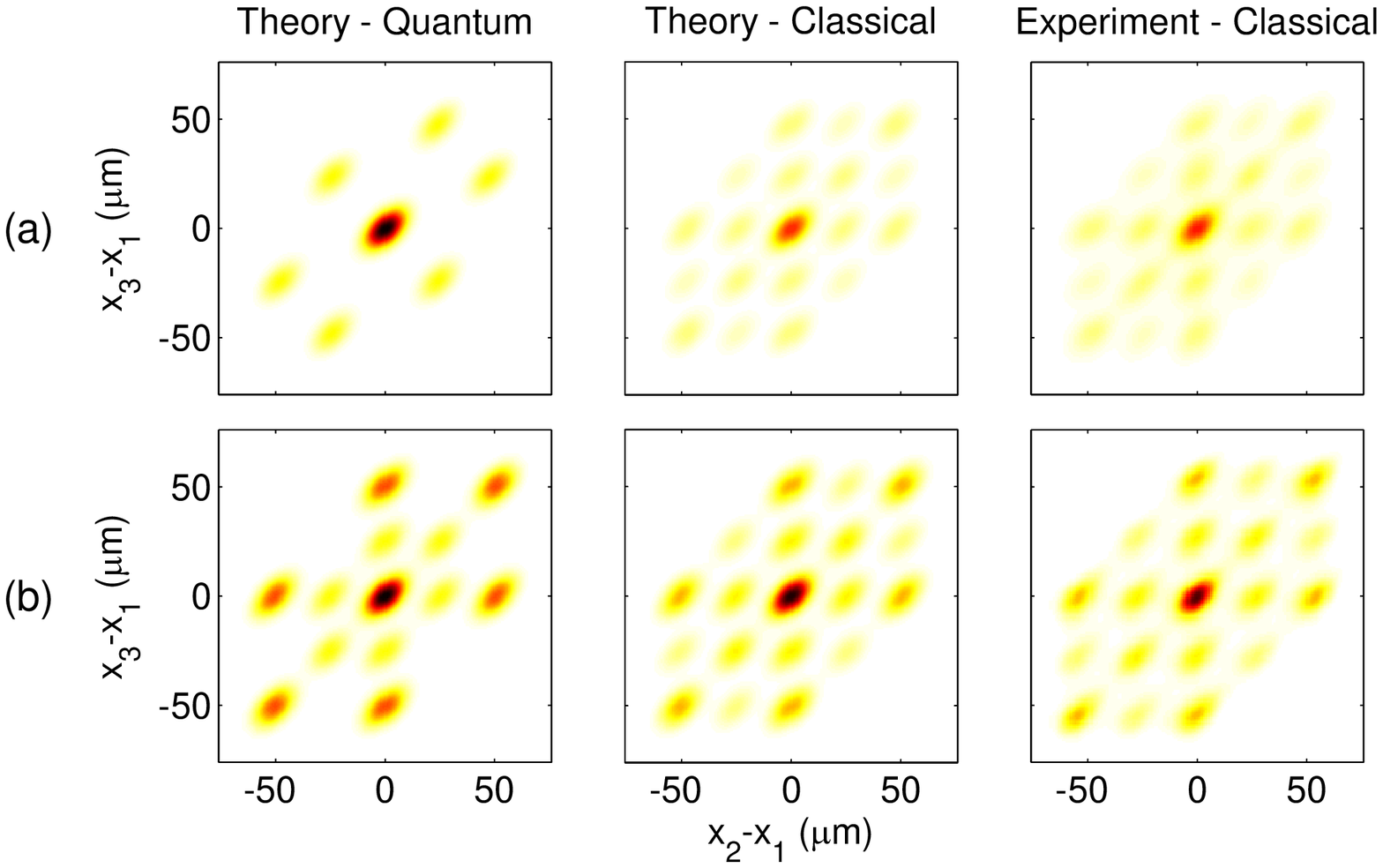}
  \caption{Spatial third order intensity correlation functions for three beams entering three-way beam splitters. (a) [(b)] An equal (unequal) beam splitter, $D$=$70 \mu m$ [$D$=$74 \mu m$]. The x and y axes present position differences between photons.}
  \label{fig:4}
\end{figure}

Fig.~\ref{fig:4} presents third-order correlation maps for three beams entering a waveguide set as either one of the two types of three-way beam-splitters discussed above, at the locations $-D/3$, 0 and $D/3$. Fig.~\ref{fig:4}(a) [(b)] presents the case of the equal [unequal] three-way beam splitter. In order to conveniently present the third-order correlation functions, their dimensionality was reduced from three to two.  This was done by representing them as functions of position \emph{differences} between photons instead as of absolute photon positions, by taking the average of all elements with the same differences. Three-photon bunching is represented by the middle spot. Two photon bunching is represented by the spots on the two main axes, and on the main diagonal. All other spots represent complete anti-bunching. The measurement for thermal light (right-most panels) is compared against the calculations for thermal (middle panels) and quantum (left panels) input states. The correspondence between experiment and theory is clearly seen.
%
%
%

In summary, it is shown that the ratio between the recurrence period of the correlations between photons propagating through a multi-mode waveguide and the recurrence period of their total intensity distribution depends on the number of input beams and on the symmetries of the input state. While for a totally asymmetric input state this ratio is 1, it can increase up to 2N for N equally spaced, uncorrelated, identical input beams. Cases with identical intensity distributions can thus display different correlation patterns. This is shown theoretically for quantum light, and both theoretically and experimentally for thermal light. Multi-mode waveguides could thus offer a simple yet versatile tool for the engineering and study of nontrivial states of light. 

We would like to thank Yoav Lahini for most valuable discussions. The financial support of the Minerva Foundation, the European Research Council, and the Crown Photonics Center is gratefully acknowledged.


\begin{thebibliography}{15}
\expandafter\ifx\csname natexlab\endcsname\relax\def\natexlab#1{#1}\fi
\expandafter\ifx\csname bibnamefont\endcsname\relax
  \def\bibnamefont#1{#1}\fi
\expandafter\ifx\csname bibfnamefont\endcsname\relax
  \def\bibfnamefont#1{#1}\fi
\expandafter\ifx\csname citenamefont\endcsname\relax
  \def\citenamefont#1{#1}\fi
\expandafter\ifx\csname url\endcsname\relax
  \def\url#1{\texttt{#1}}\fi
\expandafter\ifx\csname urlprefix\endcsname\relax\def\urlprefix{URL }\fi
\providecommand{\bibinfo}[2]{#2}
\providecommand{\eprint}[2][]{\url{#2}}

\bibitem[{\citenamefont{Robinett}(2004)}]{Robinett04}
\bibinfo{author}{\bibfnamefont{R.~W.} \bibnamefont{Robinett}},
  \bibinfo{journal}{Phys. Rep.} \textbf{\bibinfo{volume}{392}},
  \bibinfo{pages}{1} (\bibinfo{year}{2004}).

\bibitem[{\citenamefont{Yeazell et~al.}(1990)\citenamefont{Yeazell, Mallalieu,
  and \mbox{Stroud Jr.}}}]{Yeazell90}
\bibinfo{author}{\bibfnamefont{J.~A.} \bibnamefont{Yeazell}},
  \bibinfo{author}{\bibfnamefont{M.}~\bibnamefont{Mallalieu}},
  \bibnamefont{and} \bibinfo{author}{\bibfnamefont{C.~R.}
  \bibnamefont{\mbox{Stroud Jr.}}}, \bibinfo{journal}{Phys. Rev. Lett.}
  \textbf{\bibinfo{volume}{64}}, \bibinfo{pages}{2007} (\bibinfo{year}{1990}).

\bibitem[{\citenamefont{Clauser and Li}(1994)}]{Clauser94}
\bibinfo{author}{\bibfnamefont{J.~F.} \bibnamefont{Clauser}}, 
  \bibinfo{author}{\bibfnamefont{S.~F.} \bibnamefont{Li}},
  \bibinfo{journal}{Phys. Rev. A} \textbf{\bibinfo{volume}{49}},
  \bibinfo{pages}{R2213} (\bibinfo{year}{1994}).

\bibitem[{\citenamefont{Chapman et~al.}(1995)}]{Chapman95}
\bibinfo{author}{\bibfnamefont{M.~S.} \bibnamefont{Chapman}}
  \bibnamefont{et~al.}, \bibinfo{journal}{Phys. Rev. A}
  \textbf{\bibinfo{volume}{51}}, \bibinfo{pages}{R14} (\bibinfo{year}{1995}).

\bibitem[{\citenamefont{Brezger et~al.}(2002)}]{Brezger02}
\bibinfo{author}{\bibfnamefont{B.}~\bibnamefont{Brezger}} \bibnamefont{et~al.},
  \bibinfo{journal}{Phys. Rev. Lett.} \textbf{\bibinfo{volume}{88}},
  \bibinfo{pages}{100404} (\bibinfo{year}{2002}).

\bibitem[{\citenamefont{Deng et~al.}(1999)}]{Deng99}
\bibinfo{author}{\bibfnamefont{L.}~\bibnamefont{Deng}} \bibnamefont{et~al.},
  \bibinfo{journal}{Phys. Rev. Lett.} \textbf{\bibinfo{volume}{83}},
  \bibinfo{pages}{5407} (\bibinfo{year}{1999}).

\bibitem[{\citenamefont{Talbot}(1836)}]{Talbot}
\bibinfo{author}{\bibfnamefont{H.~F.} \bibnamefont{Talbot}},
  \bibinfo{journal}{Philos. Mag.} \textbf{\bibinfo{volume}{9}},
  \bibinfo{pages}{401} (\bibinfo{year}{1836}).

\bibitem[{\citenamefont{Rayleigh}(1881)}]{Rayleigh}
\bibinfo{author}{\bibfnamefont{L.}~\bibnamefont{Rayleigh}},
  \bibinfo{journal}{Philos. Mag.} \textbf{\bibinfo{volume}{11}},
  \bibinfo{pages}{196} (\bibinfo{year}{1881}).

\bibitem[{\citenamefont{Soldano and Pennings}(1995)}]{Soldano95}
\bibinfo{author}{\bibfnamefont{L.~B.} \bibnamefont{Soldano}} \bibnamefont{and}
  \bibinfo{author}{\bibfnamefont{E.~C.~M.} \bibnamefont{Pennings}},
  \bibinfo{journal}{J. Lightw. Technol.} \textbf{\bibinfo{volume}{13}},
  \bibinfo{pages}{615} (\bibinfo{year}{1995}).

\bibitem[{\citenamefont{Heaton and Jenkins}(1999)}]{Heaton99}
\bibinfo{author}{\bibfnamefont{J.~M.} \bibnamefont{Heaton}} \bibnamefont{and}
  \bibinfo{author}{\bibfnamefont{R.~M.} \bibnamefont{Jenkins}},
  \bibinfo{journal}{IEEE Photon. Technol. Lett.} \textbf{\bibinfo{volume}{11}},
  \bibinfo{pages}{212} (\bibinfo{year}{1999}).

\bibitem[{\citenamefont{Peruzzo et~al.}(2011)}]{Peruzzo11}
\bibinfo{author}{\bibfnamefont{A.}~\bibnamefont{Peruzzo}} \bibnamefont{et~al.},
  \bibinfo{journal}{Nature Commun.} \textbf{\bibinfo{volume}{2}},
  \bibinfo{pages}{224} (\bibinfo{year}{2011}).

\bibitem[{\citenamefont{\mbox{Hanbury~Brown} and Twiss}(1956)}]{HBT}
\bibinfo{author}{\bibfnamefont{R.}~\bibnamefont{\mbox{Hanbury~Brown}}}
  \bibnamefont{and}
  \bibinfo{author}{\bibfnamefont{R.~Q.} \bibnamefont{Twiss}},
  \bibinfo{journal}{Nature} \textbf{\bibinfo{volume}{178}},
  \bibinfo{pages}{1046} (\bibinfo{year}{1956}).

\bibitem[{\citenamefont{Hong et~al.}(1987)\citenamefont{Hong, Ou, and
  Mandel}}]{HOM87}
\bibinfo{author}{\bibfnamefont{C.~K.} \bibnamefont{Hong}},
  \bibinfo{author}{\bibfnamefont{Z.~Y.} \bibnamefont{Ou}}, \bibnamefont{and}
  \bibinfo{author}{\bibfnamefont{L.}~\bibnamefont{Mandel}},
  \bibinfo{journal}{Phys. Rev. Lett.} \textbf{\bibinfo{volume}{59}},
  \bibinfo{pages}{2044} (\bibinfo{year}{1987}).

%
%
\bibitem[{\citenamefont{Lou et~al.}(2009)}]{Lou09}
\bibinfo{author}{\bibfnamefont{K.~H.} \bibnamefont{Lou}} \bibnamefont{et~al.},
  \bibinfo{journal}{Phys. Rev. A} \textbf{\bibinfo{volume}{80}},
  \bibinfo{pages}{043820} (\bibinfo{year}{2009});
  \bibinfo{author}{\bibfnamefont{K.~H.} \bibnamefont{Lou}} \bibnamefont{et~al.},
  \bibinfo{journal}{Phys. Rev. A} \textbf{\bibinfo{volume}{83}},
  \bibinfo{pages}{029902(E)} (\bibinfo{year}{2011}).

\bibitem[{\citenamefont{Song et~al.}(2010)}]{Song10}
\bibinfo{author}{\bibfnamefont{X.~B.} \bibnamefont{Song}} \bibnamefont{et~al.},
 \bibinfo{journal}{Phys. Rev. A} \textbf{\bibinfo{volume}{82}},
  \bibinfo{pages}{033823} (\bibinfo{year}{2010}).

\bibitem[{\citenamefont{Song et~al.}(2011)}]{Song11}
\bibinfo{author}{\bibfnamefont{X.~B.} \bibnamefont{Song}} \bibnamefont{et~al.},
  \bibinfo{journal}{Phys. Rev. Lett.} \textbf{\bibinfo{volume}{107}},
  \bibinfo{pages}{033902} (\bibinfo{year}{2011}).

\bibitem[{\citenamefont{Bromberg et~al.}(2009)}]{Bromberg09}
\bibinfo{author}{\bibfnamefont{Y.}~\bibnamefont{Bromberg}}
  \bibnamefont{et~al.}, \bibinfo{journal}{Phys. Rev. Lett.}
  \textbf{\bibinfo{volume}{102}}, \bibinfo{pages}{253904}
  (\bibinfo{year}{2009}).

\bibitem[{\citenamefont{Tichy et~al.}(2010)}]{Tichy10}
\bibinfo{author}{\bibfnamefont{M.~C.} \bibnamefont{Tichy}}
  \bibnamefont{et~al.}, \bibinfo{journal}{Phys. Rev. Lett.}
  \textbf{\bibinfo{volume}{104}}, \bibinfo{pages}{220405}
  (\bibinfo{year}{2010}).

\end{thebibliography}
\end{document}